\documentclass[preprint]{elsarticle}

\usepackage{euscript,amssymb,amsmath}

\usepackage{graphicx}
\usepackage{epsfig}

\newcommand{\be}[1]{\begin{equation}\label{#1}}
\newcommand{\ee}{\end{equation}}
\newcommand{\ba}[1]{\begin{eqnarray}\label{#1}}
\newcommand{\ea}{\end{eqnarray}}
\newcommand{\rf}[1]{(\ref{#1})}
\newcommand{\nn}{\nonumber}

\begin{document}

\begin{frontmatter}

\title{Significance of tension for gravitating masses\\ in Kaluza-Klein models}

\author{Maxim Eingorn}
\ead{maxim.eingorn@gmail.com}

\author{Alexander Zhuk}
\ead{ai.zhuk2@gmail.com}

\address{Astronomical Observatory and Department of Theoretical Physics,\\ Odessa National University, Street Dvoryanskaya 2, Odessa 65082, Ukraine}

\begin{abstract}
In this letter, we consider the six-dimensional Kaluza-Klein models with spherical compactification of the internal space. Here, we investigate the case of bare
gravitating compact objects with the dustlike equation of state $\hat p_0=0$ in the external (our) space and an arbitrary equation of state $\hat p_1=\Omega \hat
\varepsilon$ in the internal space, where $\hat \varepsilon$ is the energy density of the source. This gravitating mass is spherically symmetric in the external space
and uniformly smeared over the internal space. In the weak field approximation, the conformal variations of the internal space volume generate the admixture of the
Yukawa potential to the usual Newton's gravitational potential. For sufficiently large Yukawa masses, such admixture is negligible and the metric coefficients of the
external spacetime coincide with the corresponding expressions of General Relativity. Then, these models satisfy the classical gravitational tests. However, we show that
gravitating masses acquire effective relativistic pressure in the external space. Such pressure contradicts the observations of compact astrophysical objects (e.g., the
Sun). The equality $\Omega =-1/2$ (i.e. tension) is the only possibility to preserve the dustlike equation of state in the external space. Therefore, in spite of
agreement with the gravitational experiments for an arbitrary value of $\Omega$, tension ($\Omega=-1/2$) plays a crucial role for the considered models.
\end{abstract}

\begin{keyword} extra dimensions \sep Kaluza-Klein models \sep tension \sep gravitational tests \end{keyword}

\end{frontmatter}


\section{\label{sec:1}Introduction}

\setcounter{equation}{0}

The multidimensionality of spacetime is an essential property of the modern theories of unification such as superstrings, supergravity and M-theory, which have the most
self-consistent formulation in spacetime with extra dimensions \cite{Polchinski}. Obviously, these physical theories should be consistent with observations. For example,
in the weak field limit they must satisfy gravitational experiments such as the perihelion shift, the deflection of light, the time delay of radar echoes and
parameterized post-Newtonian parameters. It is well known that General Relativity
is in good agreement with these experiments \cite{Will}.
Therefore, to investigate the similar correspondence for multidimensional theories,  in our papers \cite{EZ3,EZ4,EZ5} we have considered popular Kaluza-Klein models with
toroidal compactification of the internal space. We have shown that a matter source in the form of a dustlike compact gravitating object failed with the observations.
Here, the dustlike equation of state $p=0$ holds in all spatial dimensions. The obtained result was surprising to us because this approach is the most natural one for
the ordinary astrophysical objects (such as our Sun) and it works well in General Relativity \cite{Landau}. It turned out that to satisfy the experimental data, the
matter source must have negative pressure (i.e. tension\footnote{For black strings and black branes, the notion of tension is defined, e.g., in \cite{TF} and it follows
from the first law for black hole spacetimes \cite{TZ,HO,TK}.})
in the internal spaces \cite{EZ4,EZ5}. We have shown that latent solitons (in particular, the uniform black strings and black branes)
satisfy the gravitational experiments at the same level of accuracy as General Relativity. In general case, the variation of the total volume of the internal spaces
generates the fifth force \cite{EZ6}. This is the main reason of the problem. However, in the case of the latent solitons, tension of the gravitating source fixes the
internal space volume, eliminating the fifth force contribution and resulting in agreement with the observations. Therefore, tension plays a crucial role here. For
uniform black strings/branes with toroidal compactification, the equation of state in the internal spaces is $\hat p_1=-\hat \varepsilon/2$, where $\hat \varepsilon$ is
the energy density of the gravitating source. The problematic aspect of these models consists in physically reasonable explanation of the origin of tension for ordinary
astrophysical objects.

Therefore, in \cite{ChEZ1,ChEZ2} we considered a dustlike (in all spatial dimensions, i.e. without tension in the extra dimensions) matter source for Kaluza-Klein models
with spherical compactification of the internal space. In contrast to the case of toroidal compactification, this model can satisfy the gravitational experiments if the
internal space is stabilized what happens for a positive six-dimensional cosmological constant \cite{ChEZ2,ChEZ3}. Here, the fifth force is replaced by Yukawa
interaction which is short-range for large Yukawa masses (i.e. the large mass of radion). Therefore, at large three-dimensional distances, the effect of this interaction
is negligibly small. Roughly speaking, the agreement with observations occurs asymptotically. Moreover, all models where a matter source, i.e. a compact gravitating
massive body with the energy density $\hat \varepsilon$, has the dustlike equation of state $\hat p_0=0$ in the external (our) space and an arbitrary equation of state
$\hat p_1=\Omega \hat \varepsilon$ in the internal space satisfy asymptotically (in the region of negligibly small Yukawa interaction) the gravitational experiments
\cite{ChEZ3}. With increase in the distance $r_3$ (the length of the radius-vector in the three-dimensional space) from a massive source, all these models tend
asymptotically to the exact black brane solution with spherical compactification \cite{ChEZ3}. For such exact black brane solution, the parameter of state $\Omega
=-1/2$, in full analogy with black branes with toroidal compactification. Therefore, for any $\Omega$ (including the dustlike value $\Omega=0$), considered models can
satisfy the gravitational experiments. Honestly speaking, this result was not surprising to us because it confirms the conventional wisdom that stabilization of the
internal space solves the problem with observations. However, the novelty of our paper is that stabilization is not sufficient. The gravitational tests are not the only
possible experiments. For example, gravitating bodies such as a system of nonrelativistic particles must also have certain thermodynamical properties. The analysis
carried out in the present paper shows that in all models with $\Omega \neq -1/2$ gravitating matter sources, e.g. nonrelativistic particles, acquire effective
relativistic pressure in the external (our) space. Hence, a system of such particles will also have effective relativistic pressure, that is nonsense from the
thermodynamical point of view.
Therefore, in spite of the agreement (asymptotical) with the gravitational experiments, such models fail with the observations. This important point was not elucidated
in our previous papers. Only in the case of tension $\Omega=-1/2$, a matter source remains dustlike in the external space. Therefore, tension also plays a crucial role
in models with spherical compactification as in the case of toroidal compactification. The only problem is to explain the physical origin of tension for ordinary
astrophysical objects.

In Sec. 2, we demonstrate that a compact gravitating source in the considered models acquires effective relativistic pressure in the external space except the case of
tension in the internal space. This is the only possibility to preserve the dustlike equation of state in the external space. The main results are briefly summarized in
concluding Sec. 3.

\section{Effective energy density and pressure of the gravitating mass}

As we pointed out in papers \cite{ChEZ2,ChEZ3}, the matter source in the Kaluza-Klein models with spherical compactification should consist of two parts. First, it is
the homogeneous perfect fluid which provides spherical compactification of the internal space. Second, it is the gravitating object, which is spherically symmetric and
compact (i.e. pointlike) in the external space and uniformly smeared over the internal space. The total energy-momentum tensor is the sum of these parts with the
following nonzero components:
\ba{1}
&{}&T^0_{0}\approx\bar{\varepsilon}+\varepsilon^1+\hat \rho({\bf r}_3)c^2\, ,\\
&{}&T_{\alpha}^{\alpha}\approx\bar{\varepsilon}+\varepsilon^1\, ,\quad\alpha=1,2,3\, , \label{2}\\
&{}&T_{4}^4=T_{5}^5\approx-\omega_1\bar{\varepsilon}-\omega_1\varepsilon^1 - \Omega\hat \rho({\bf r}_3)c^2\label{3}\, ,
\ea
where $\bar{\varepsilon}$ is the energy density of the homogeneous perfect fluid, $\hat \rho({\bf r}_3)$ is the rest mass density of a compact gravitating object, ${\bf
r}_3$ is the three-dimensional radius-vector and $\varepsilon^1$ is the excitation of the background matter energy density by this object\footnote{The subscripts 0 and 1
are reserved for the parameters of the model relating to external (our) and internal spaces, respectively. The superscript 1 here and below marks the perturbations. The
excitation $\varepsilon^1$ is of the same order of magnitude as $\hat \rho c^2$.}. The background matter is fine-tuned with the radius $a$ of the two-sphere:
$\bar{\varepsilon}=[(1+\omega_1)\kappa_6 a^2]^{-1}$, and a free parameter $\omega_1$ defines the equation of state of this matter in the internal space. The model may
also include a six-dimensional cosmological constant $\Lambda_6$, which is fine-tuned with the parameters of the model: $\Lambda_6=\omega_1\bar{\varepsilon}$. This bare
cosmological constant is absent if $\omega_1=0$. The gravitating compact object has the dustlike equation of state in the external (our) space $\hat p_0 =0$ and an
arbitrary equation of state $\hat p_1 \approx \Omega \hat \rho({\bf r}_3)c^2$ in the internal space. We also suppose that this object is uniformly smeared over the
internal space: $\hat \rho({\bf r}_3)=\hat \rho_3({\bf r}_3)/V_2$ where $V_2=4\pi a^2$. In the case of a pointlike mass in the external space $\hat \rho_3({\bf
r}_3)=m\delta({\bf r}_3)$. It also worth noting that the six-dimensional and Newton's gravitational constants are related as follows: $\kappa_6/V_2=\kappa_N \equiv 8\pi
G_N/c^4$.

The metrics for the considered model in isotropic coordinates takes the form (see for details \cite{ChEZ1,ChEZ2})
$$
ds^2=Ac^2dt^2+Bdx^2+Cdy^2+Ddz^2+E(d\xi^2+\sin^2\xi d\eta^2)
$$
with $A \approx 1+A^{1}({r}_3)$, $B \approx -1+B^{1}({r}_3)$, $C \approx -1+C^{1}({r}_3)$, $D \approx -1+D^{1}({r}_3)$, $E \approx -a^2+E^{1}({r}_3)$, where all metric
perturbations $A^1,B^1,C^1,D^1,E^1$ are of the order $O(1/c^2)$ and can be found with the help of the Einstein equations. They read
\ba{4}
A^1&=&\frac{2\varphi_N}{c^2}+\frac{E^1}{a^2}\, , \\
\label{5} B^1&=&C^1=D^1=\frac{2\varphi_N}{c^2}-\frac{E^1}{a^2}\, ,\\
\label{6}
E^1&=&a^2\cfrac{\varphi_N}{c^2}\,\left( 1+2\Omega  \right) e^{-r_3/\lambda}, \quad \lambda\equiv a/\sqrt{\omega_1}\, ,
\ea
where the Newton's potential is $\varphi_N=-G_N m/r_3$. The solution \rf{6} takes place for $\omega_1>0$. In the opposite case $\omega_1<0$, we get the nonphysical
oscillating solution. If $\Omega \neq -1/2$, Eq. \rf{6} demonstrates that conformal variations of the internal space volume generate the Yukawa interaction.
The admixture of such interaction to $A^1,B^1,C^1,D^1$ is negligible at distances $r_3\gg \lambda$ (i.e. for the large Yukawa mass $\sqrt{\omega_1}/a$), and we achieve
good agreement with the gravitational tests in this region. Exactly this situation takes place in the Solar system \cite{ChEZ2}.

The Einstein equations also lead to the following important relation: $\varepsilon^1 = E^1/\left(\kappa_6 a^4\right)$. Eq. \rf{6} shows that this background perturbation
is localized around the gravitating mass and falls off exponentially with the distance $r_3$ from the gravitating object. Therefore, the bare gravitating mass is covered
by this "coat".  For an external observer, this coated gravitating mass is characterized by the effective energy-momentum tensor with the following nonzero components:
\ba{8}
&{}&T_{0}^{0(eff)}\approx \varepsilon^1+\hat\rho({\bf r}_3)c^2\nn\\
&{}&=-(1+2\Omega)\frac{mc^2}{2V_2^2r_3}\exp\left(-\frac{\sqrt{\omega_1}}{a}r_3\right)+\frac{1}{V_2}mc^2\delta({\bf r}_3),\\
&{}&T_{\alpha}^{\alpha (eff)}\approx \varepsilon^1\label{9}\nn\\
&{}&=-(1+2\Omega)\frac{mc^2}{2V_2^2r_3}\exp\left(-\frac{\sqrt{\omega_1}}{a}r_3\right),\quad\alpha=1,2,3\, ,\\
&{}&T_{4}^{4(eff)}=T_{5}^{5(eff)}\approx-\omega_1\varepsilon^1-\Omega\hat\rho({\bf r}_3) c^2\label{10}\nn\\
&{}&=(1+2\Omega)\frac{\omega_1mc^2}{2V_2^2r_3}\exp\left(-\frac{\sqrt{\omega_1}}{a}r_3\right)-\frac{\Omega}{V_2}mc^2\delta({\bf r}_3)\, .
\ea
These components define the effective energy density and pressure of the coated gravitating mass. For example, from Eq. \rf{9} we conclude that this mass acquires
relativistic pressure $\hat p_{0}^{(eff)}=-T_{\alpha}^{\alpha (eff)}$ in the external space. To demonstrate it more clearly, we can replace the rapidly decreasing
exponential function by the delta function:
\ba{11} \frac{1}{r_3}\exp\left(-\frac{\sqrt{\omega_1}}{a}r_3\right)\rightarrow\int\frac{1}{r'_3}\exp\left(-\frac{\sqrt{\omega_1}}{a}r'_3\right)dV'_3 \times \delta({\bf
r}_3)=\frac{V_2}{\omega_1} \delta({\bf r}_3)\, . \ea
Then, Eqs. \rf{8}-\rf{10} read
\ba{12}
&{}&T_{0}^{0(eff)}\rightarrow-(1+2\Omega)\frac{mc^2}{2V_2\omega_1}\delta({\bf r}_3)+\frac{1}{V_2}mc^2\delta({\bf r}_3)\nn \\
&{}&=\hat \rho({\bf r}_3)c^2\left(1-\frac{1+2\Omega}{2\omega_1}\right)\, ,\\
&{}&T_{\alpha}^{\alpha(eff)}\rightarrow -(1+2\Omega)\frac{mc^2}{2V_2\omega_1}\delta({\bf r}_3)\nn \\
&{}&=-\hat \rho({\bf r}_3)c^2\frac{1+2\Omega}{2\omega_1}\label{13},\quad\alpha=1,2,3\, ,\\
&{}&T_{4}^{4(eff)}\rightarrow(1+2\Omega)\frac{mc^2}{2V_2}\delta({\bf r}_3)-\frac{\Omega}{V_2}mc^2\delta({\bf r}_3)\nn \\
&{}&=\hat \rho({\bf r}_3)c^2\frac{1}{2}\label{14}\, .
\ea
These equations give us the effective energy density and pressure of the coated gravitating mass. We see that the effective energy density $\hat \varepsilon^{(eff)}
=T_{0}^{0(eff)}$ and effective pressure in the external (our) space $\hat p_{0}^{(eff)}=-T_{\alpha}^{\alpha (eff)}$ depend on parameter $\Omega$, which defines the
equation of state of the bare gravitating mass in the internal space. On the other hand, the effective pressure in the internal space $\hat p_{1}^{(eff)}=-T_{4}^{4
(eff)}$ does not depend on $\Omega$ and is negative. From Eq. \rf{13}, we clearly see that the coated gravitational mass acquires relativistic pressure in the external
(our) space. Obviously, it is not the case for compact astrophysical objects, such as our Sun. Usually, they have nonrelativistic velocities in the three-dimensional
space, and their pressure is much less than the energy density. Moreover, in the weak field limit the pressure for these objects is taken in the dustlike form $\hat
p_0=0$. It can be easily seen that the equality $\Omega =-1/2$ is the only possibility to achieve $\hat p_{0}^{(eff)}=0$ for our model\footnote{In the case of a very
large parameter of the equation of state of background matter in the internal space $\omega_1 \gg 1$, the effective pressure $\hat p_0^{(eff)}$ can be also
nonrelativistic. However, such matter is very unrealistic. We are not aware of any example of it. In realistic cases $\omega_1\sim 1$. For example, $\omega_1=1$
corresponds to the monopole form-fields (the Freund-Rubin scheme of compactification), and for the Casimir effect we have $\omega_1=2$ \cite{exci,Zhuk}.}. It means that
the bare gravitating mass should have tension with equation of state $\hat p_1=-\hat \varepsilon /2$ in the internal space. Then, the effective and bare energy densities
coincide with each other and the gravitating mass remains pressureless in our space. In the internal space the gravitating mass still has tension with the parameter of
state $-1/2$. Therefore, tension plays a crucial role for models with spherical compactification.

\section{Conclusion}

In this letter, we have considered the six-dimensional Kaluza-Klein models with spherical compactification of the internal space.
A bare gravitating mass has the dustlike equation of state $\hat p_0=0$ in the external (our)
space and an arbitrary equation of state $\hat p_1=\Omega \hat \varepsilon$ in the internal space, and
it disturbs the background matter which provides spherical compactification of the internal space. This perturbation takes the form of the Yukawa potential and is
localized around the bare mass. For any value of $\Omega$, the Yukawa interaction is short-ranged for the large mass of radion, providing agreement with the
gravitational experiments in the corresponding asymptotic region \cite{ChEZ2,ChEZ3}. However, the gravitational tests are not the only possible experiments. For example,
gravitating bodies, such as a system of nonrelativistic particles, must have certain thermodynamical properties.
In the present paper we have shown that due to localized perturbations, gravitating massive sources are covered by this coat for models with $\Omega\neq -1/2$. As a
result, these coated gravitating masses acquire effective relativistic pressure in the external space. Hence, a system of nonrelativistic particles will also have
effective relativistic pressure, which is unacceptable from the thermodynamical point of view. Therefore, in spite of the agreement with the gravitational experiments,
such models fail with the observations. The equality $\Omega =-1/2$ is the only possibility to preserve the dustlike equation of state in the external space. Therefore,
to be in agreement with observations (both gravitational and thermodynamical), bare gravitating masses in multidimensional Kaluza-Klein models with spherical
compactification should have tension with $\Omega=-1/2$ in the internal space. This important point was not elucidated in our previous papers.

It worth noting that in
five-dimensional models with toroidal compactification, the need for tension $-1/2$ for compact astrophysical objects was indicated in the pioneering paper \cite{ChD}.
This follows from the gravitational tests in the weak field limit. However, the authors wrote that an argument why these objects should have such tension is absent.
Unfortunately, such physically reasonable argument is absent up to now.


\section*{Acknowledgements}

This work was supported in part by the "Cosmomicrophysics-2" programme of the Physics and Astronomy Division of the National Academy of Sciences of Ukraine.


\end{document}